\documentclass[%
 reprint,
 amsmath,amssymb,
 aps,
]{revtex4-2}

\usepackage{graphicx}% Include figure files
\usepackage{dcolumn}% Align table columns on decimal point
\usepackage{bm}% bold math
\usepackage{physics}
\usepackage{amsmath}
\usepackage{amssymb}
\usepackage{subcaption}
\usepackage{xcolor}
\graphicspath{{./image/}}
\begin{document}

\preprint{APS/123-QED}

\title{An analytical theory of CEP-dependent coherence driven by few-cycle pulses}

\author{Bing Zeng}
 \affiliation{Department of Physics and Astronomy, The University of Alabama in Huntsville, Huntsville, AL 35899, USA.}
\author{Lingze Duan}
\email{lingze.duan@uah.edu}
\affiliation{Department of Physics and Astronomy, The University of Alabama in Huntsville, Huntsville, AL 35899, USA.}


%
\date{\today}% It is always \today, today,

\begin{abstract}
The interaction between an atomic system and a few-cycle ultrafast pulse carries rich physics and a considerable application prospect in quantum-coherence control. However, theoretical understanding of its general behaviors has been hindered by the lack of an analytical description in this regime, especially with regard to the impact of the carrier-envelope phase (CEP). Here, we present an analytical theory that describes a two-level atom driven by a far-off-resonance, few-cycle square pulse. A simple, closed-form solution of the Schrödinger equation is obtained under the first-order perturbation without invoking the rotating-wave approximation or the slowly varying envelope approximation. Further investigation reveals an arithmetic relation between the final inversion of the atom and the CEP of the pulse. Despite its mathematical simplicity, the relation is able to capture some of the key features of the interaction, which prove to be robust against generalization of pulse shapes and show good agreements with numerical solutions. The theory can potentially offer a general guidance in future studies of CEP-sensitive quantum coherence.
\begin{description}
\item[DOI]
XXX

\end{description}
\end{abstract}

\maketitle

%\tableofcontents

\section{\label{sec:level1}Introduction}

Light-matter interactions have drawn a lot of interest in quantum optics and atomic physics in recent years. Driven by the growing prospect of quantum information, considerable effort has been dedicated in the research of quantum coherence and quantum interference in atomic systems \cite{Ficek2004,Harris1998,Bloch2008}. Over the last two decades, the advance in ultrafast laser technology has made few-cycle pulses widely available \cite{Brabec2000,Kartner2004}. This enables the study of quantum coherence in the few-cycle regime, which leads to a number of new discoveries \cite{Ziolkowski1995,Cheng2001,Wu2007} and applications \cite{Vlack2007,Kumar2013,Yang2014}. 

Few-cycle excitation of a two-level atom bears some unique features. Generally, the area theorem is no longer valid for few-cycle pulses, and the actual field pattern underneath the pulse envelope has a direct impact on the interaction outcome \cite{Ziolkowski1995,Hughes1998}. The field pattern in a few-cycle pulse is often characterized by the so-called carrier-envelope phase (CEP), the phase of the optical carrier relative to the peak of the pulse envelope. When the pulse width becomes comparable to the optical cycle, CEP can strongly affect the shape of the oscillating field \cite{Brabec2000}. As a result, CEP has proved to be a critical factor in strong-field interactions such as high-harmonic generation \cite{Christov1997}, above-threshold ionization \cite{Paulus2001} and attosecond pulse generation \cite{Sansone2006}. Moreover, due to the breakdown of the rotating-wave approximation (RWA) and the slowly varying envelope approximation (SVEA) in the few-cycle regime \cite{Kalosha1999,Jirauschek2005,Leblond2013}, Bloch equations without RWA and SVEA have to be adopted, which increases the complexity of the problem significantly. Consequently, theoretical analysis in this regime has been primarily based on numerical approaches.

The lack of an analytical description has rendered the physics of few-cycle interactions less clear. Most of the findings so far have been based on numerical solutions aligned towards specific problems, while a general picture of the qualitative trend is missing. There has been some prior effort in pursuit of analytical solutions in the few-cycle regime \cite{Roudnev2007,Rostovtsev2009,Begzjav2017}. One particularly interesting area is the study of atomic coherence driven by far-off-resonance pulses \cite{Rostovtsev2009}. Far-off-resonance pulses are of practical significance because of their potential applications in efficient generation of soft x-ray and UV radiations \cite{Scully2008,Jha2012}. A suite of analytical solutions have been obtained under the RWA for far-detuned many-cycle pulses \cite{Jha2010,Jha20102}. In the few-cycle regime, however, the perturbation theory has to be invoked to simplify the Schrödinger equation under strong excitation, which yields a partial solution in the form of a nested integral \cite{Rostovtsev2009,Jha2012}. This solution is implicit and would still require numerical methods when solving actual problems.

In this work, we take a different approach to solve the Schrödinger equation. By assuming a far-detuned, few-cycle square pulse, we are able to obtain a simple, closed-form solution. The solution is further applied to evaluating the impact of CEP on atomic inversion, which leads to a concise, arithmetic expression. Comparisons with direct numerical solutions of the Schrödinger equation show good agreements. Key characteristics of the CEP dependence are found to be robust against generalization of the pulse shapes, indicating the feasibility of the method in offering a sound physical picture beyond the square pulse. As such, this analytical theory appears to be able to complement the traditional numerical methods with a qualitative guideline.

\section{Theoretical model}
Our theoretical model is based on the general framework of a two-level system (TLS) under the excitation of an electromagnetic field. The Hamiltonian of the TLS is given by
%%%%%%%%%%%%%%%%%%%%%%%%%%%%%%%%%%%%%equation(1)
\begin{equation}
\hat{H}=\hbar \omega_c \ket{c} \bra{c}-\mu \mathcal{E}(t) \ket{c} \bra{d} - [\mu \mathcal{E}(t)]^*\ket{d} \bra{c},
\end{equation}

\noindent where $\ket{c}$ and  $\ket{d}$ are upper and lower level states of the TLS, respectively. $\mu$ is the dipole moment. $\omega_c$ is the transition frequency of the TLS. $\mathcal{E}(t)=E(t)\cos(\omega t + \varphi)$ is the excitation field defined in the moving frame, where $E(t)$ represents the pulse envelop, $\omega$ is the carrier wave frequency, and $\varphi$ is the CEP. The medium is assumed to be optically thin so the propagation effects are neglected. From the Schrödinger equation, the following relations can be established,
%%%%%%%%%%%%%%%%%%%%%%%%%%%%%%%%%%%%%equation(2)
\begin{subequations}
\begin{align}
\begin{split}
\Dot{C}(t)=-i \Omega (t) \cos(\omega t +\varphi) e^{i\omega_c t} D(t),
\end{split}\\
\begin{split}
\Dot{D}(t)=-i \Omega^* (t) \cos(\omega t +\varphi) e^{-i\omega_c t} C(t),
\end{split}
\end{align}

\end{subequations}

\noindent where $C(t)$ and $D(t)$ are probability amplitudes of the state vector $\ket{\Psi}$, with $\ket{\Psi}=C(t) e^{-i\omega_c t} \ket{c} + D(t) \ket{d}$. $\Omega(t)=\mu E(t)/{\hbar}$ is the Rabi frequency.

By defining $f(t) = C(t)/D(t)$, the Schrödinger equation (2) can be converted into a Riccati equation without invoking the RWA or the SVEA \cite{Rostovtsev2009},
\begin{equation}
\begin{split}
\Dot{f}  ={}&i \Omega^*(t) \cos(\omega t +\varphi) e^{-i\omega_c t} f^2\\ 
&-i \Omega(t) \cos(\omega t +\varphi) e^{i\omega_c t}.
\end{split}
\end{equation}

\noindent Eq. (3) can be simplified under the first-order perturbation by introducing $g(t)=f(t)+i\theta(t)$, where
\begin{equation}
\theta(t)=\int_{-\infty}^t \Omega(t')\cos(\omega t' +\varphi)e^{i\omega_c t'} d t'.
\end{equation}

\noindent Since $-i\theta(t)$ is the solution of (3) when the $f^2$ term on the right hand side of (3) is set to zero, $g(t)$ is effectively the difference between $f(t)$ and its zeroth-order approximation. Substituting $g(t)$ into (3) and neglecting the $g^2$ term yields the equation
\begin{equation}
\Dot{g} - 2 \theta \Dot{\theta}^* g =  -i \theta^2 \Dot{\theta}^*.
\end{equation}

\noindent Applying the Green's function to (5) leads to the general solution
\begin{equation}
g(t) = -i \int_{-\infty}^{t} \theta^2 \Dot{\theta}^* e^{\alpha(t';t)} d t',
\end{equation}

\noindent where $\alpha(t';t)$ is given by
\begin{equation}
\alpha(t';t) = 2 \int_{t'}^{t} \theta \Dot{\theta}^* d t''.
\end{equation}

\noindent Since $d\alpha/dt' = -2 \theta \Dot{\theta}^*$, (6) can be simplified to
\begin{equation}
g(t) = \frac{i}{2} \int_{-\infty}^{t} \theta(t') d e^{\alpha(t';t)}.
\end{equation}

\noindent The right-hand side of (8) can be integrated by parts using the conditions $\theta(-\infty) \approx 0$ and $\alpha(t;t) = 0$. This then leads to the general solution of $f(t)$,
\begin{equation}
f(t) = -\frac{i}{2} \left[ \theta(t) + \int_{-\infty}^{t} \Dot{\theta}(t') e^{\alpha(t';t)} d t' \right].
\end{equation}

%%%%%%%%%%%%%%%%%%%%%%%%%%%%%%%%%%%%%%%%%%%%%%%%%%%%%% End of 1st Part
Note that a slightly different version of this integration-form solution has been given in Ref. 20. The focus of the current paper, however, is to find a \textit{closed-form} solution of the Schrödinger equation for a particular type of excitation pulse, specifically, a few-cycle square pulse. The pulse is defined as
\begin{equation}
    E(t) =
\begin{cases}
    E_0, & 0 \leq t \leq \tau\\
    0, & \text{Otherwise}
\end{cases}
\end{equation}

\noindent where $\tau$ is the pulse width. Taking such a pulse into (4), it is easy to see that, within the duration of the pulse ($0 \leq t \leq \tau$), $\theta(t)$ satisfies
\begin{equation}
\theta(t) = \Omega_0 \int_{0}^{t} \cos(\omega t'+\varphi) e^{i\omega_c t'} d t',
\end{equation}

\noindent where $\Omega_0 = \mu E_0 / {\hbar}$ is the peak Rabi frequency. The integral in (11) can be separately evaluated for $\omega > \omega_c$ and $\omega < \omega_c$, and the results are similar in form. Here, we only focus on the $\omega < \omega_c$ case, with the understanding that the $\omega > \omega_c$ case can be treated in a similar fashion. A tedious but straightforward derivation shows that $\theta(t)$, $\theta^*(t)$, $\Dot{\theta}(t)$ and $\Dot{\theta}^*(t)$ can be expressed as follows:
\begin{subequations}
\begin{align}
\begin{split}
\theta = -i\eta \left[ \cos(\omega t+\varphi+i\beta) e^{i\omega_c t} - \cos(\varphi+i\beta) \right],
\end{split}\\
\begin{split}
\theta^* = i\eta \left[ \cos(\omega t+\varphi-i\beta) e^{-i\omega_c t} - \cos(\varphi-i\beta) \right],
\end{split}\\
\begin{split}
\Dot{\theta} = \Omega_0 \cos(\omega t+\varphi) e^{i\omega_c t},
\end{split}\\
\begin{split}
\Dot{\theta}^* = \Omega_0 \cos(\omega t+\varphi) e^{-i\omega_c t} ,
\end{split}
\end{align}

\end{subequations}

\noindent where $\eta = \Omega_0 / \sqrt{\omega_c^2 - \omega^2}$ and $\beta$ satisfies the relations $\cosh{\beta} = \omega_c / \sqrt{\omega_c^2 - \omega^2}$ and $\sinh{\beta} = \omega / \sqrt{\omega_c^2 - \omega^2}$. 

By substituting (12) into (7), one can show that $\alpha(t';t)$ consists of three parts written in an integration form,
\begin{equation}
\begin{split}
\alpha(t';t) = {}& - i \eta \Omega_0 \int_{t'}^{t} \{ \cos{i\beta} + \cos(2\omega t''+ 2\varphi + i\beta) \\
&- 2\cos(\varphi + i\beta) \cos(\omega t''+ \varphi) e^{-i\omega_c t''} \} d t''.
\end{split}
\end{equation}

\noindent Among these three terms, the second and the third terms have periodic amplitudes oscillating at $2\omega$ and $\omega$, respectively. When the integral covers multiple carrier cycles, the contributions of these two terms become negligible compared to the first one. Thus, we can drop the last two terms in the integral and simplify (13) to
\begin{equation}
\alpha(t';t) = -i \eta^2 \omega_c (t-t').
\end{equation}

Bringing (14) into (9), the integral in (9) can be rewritten as
\begin{equation}
\begin{split}
{}& \int_{-\infty}^{t} \Dot{\theta}(t') e^{\alpha(t';t)} d t' = \\
& e^{-i\eta^2\omega_c t} \int_{-\infty}^{t} \Omega_0 \cos(\omega t'+\varphi) e^{i(1+\eta^2)\omega_c t'} d t'.
\end{split}
\end{equation}

\noindent If $\eta$ satisfies the condition $\eta^2 \ll 1$, then $1+\eta^2 \approx 1$, and the integral on the right-hand side of (15) becomes $\theta(t)$. From (9), the final expression for $f(t)$ is
\begin{equation}
f(t) = -\frac{i}{2} \left( 1 + e^{-i\eta^2\omega_c t} \right) \theta(t),
\end{equation}

\noindent where $\theta(t)$ is given by (12a). The relation (16) is a closed-form analytical solution of the Riccati equation (3) under the first-order perturbation for square-pulse excitation. Its valid range is set by the condition $\eta^2 \ll 1$, which can be rewritten as $(\omega/\omega_c)^2 + (\Omega_0/\omega_c)^2 \ll 1$. Since $\omega < \omega_c$ is assumed in the current case, the above condition is always satisfied if we have
\begin{equation}
\left(\frac{\omega}{\omega_c}\right)^2 + \left(\frac{\Omega_0}{\omega}\right)^2 \ll 1.
\end{equation}

\noindent It is thus clear that the solution (16) is valid when the TLS is excited by a far-off-resonance square pulse with a peak Rabi frequency much less than the carrier frequency. This latter condition implies that the peak field of the pulse must stay well below the level required for the so-called \textit{carrier-wave Rabi flopping} \cite{Hughes1998,Zeng2018}. Meanwhile, the large-detuning condition also implies that $\cos{i\beta} \equiv \cosh{\beta} \approx 1$, which ensures that the simplification of (13) is valid under the assumption.

%%%%%%%%%%%%%%%%%%%%%%%Figure 1
\begin{figure}
\centering
\includegraphics[width=6.6cm, height=4.5cm]{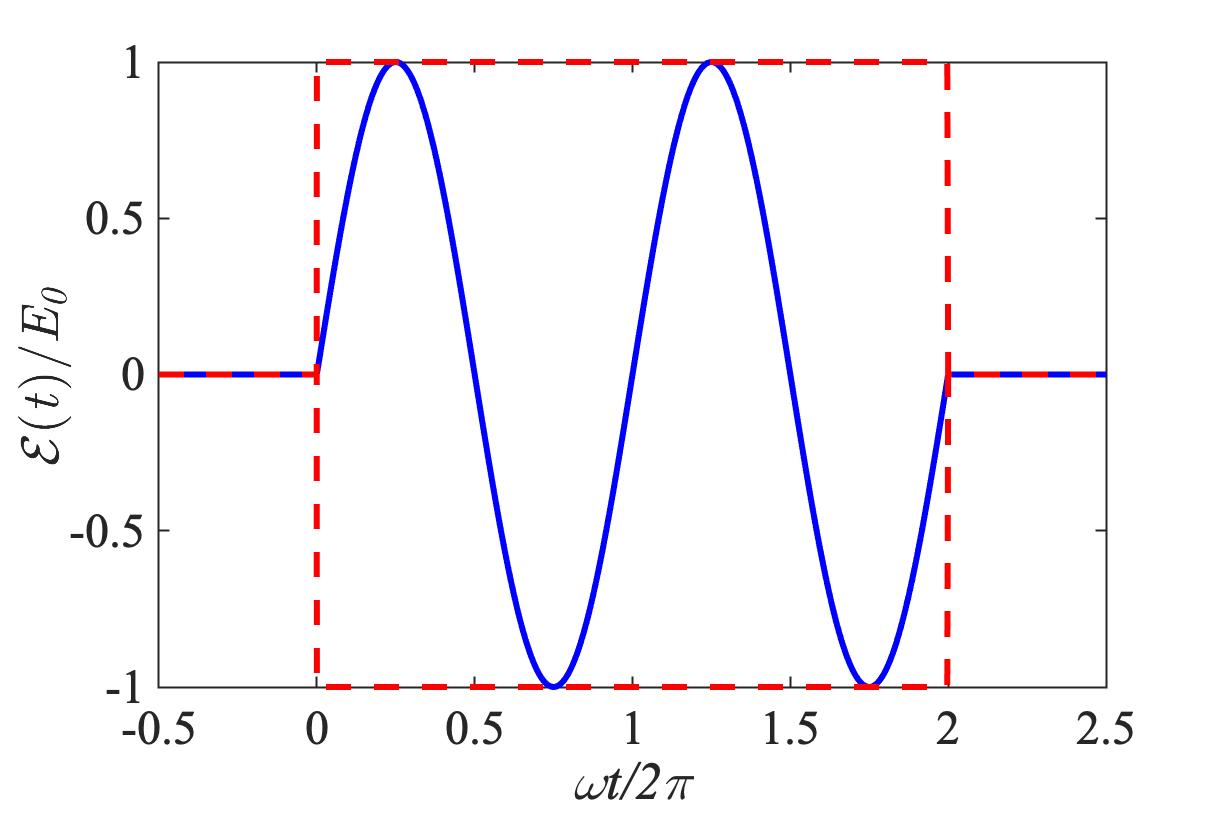} 
\caption{The incident two-cycle square pulse.}
\end{figure}

\section{Comparison with Numerical Solutions}

In order to validate the above analytical solution and also investigate its applicability, we numerically solve the Schrödinger equation (2) and compare the results with the analytical predictions made by (16). The results presented in the following are based on a two-cycle square pulse as shown in Fig. 1. Other pulse widths and field patterns have also been studied and the results are similar in nature. In all the calculations, the carrier frequency $\omega$ is fixed at a wavelength of 800 nm. The transition frequency $\omega_c$ and the Rabi frequency $\Omega_0$ are adjusted to produce different combinations of the detuning factor $\omega/\omega_c$ and the amplitude factor $\Omega_0/\omega$.

Fig. 2 shows the analytical and the numerical solutions of $|f(t)|$ across the pulse for four different peak fields ($\Omega_0/\omega =$ 0.05, 0.1, 0.15 and 0.2). In each case, three different detunings, with $\omega/\omega_c =$ 0.3, 0.6, and 0.9, are investigated. When the Rabi frequency is much lower than the carrier frequency, e.g., $\Omega_0/\omega = 0.05$, the analytical solutions show excellent agreement with the numerical results across the entire pulse for all three detuning levels. As the peak field gradually increases and the Rabi frequency moves closer towards the carrier frequency, discrepancies between the analytical and the numerical solutions begin to arise. Such differences are much more significant in the small-detuning case with $\omega/\omega_c = 0.9$.

%%%%%%%%%%%%%%%%%%%%%%%Figure 2
\begin{figure}
\centering
\includegraphics[width=8.6cm, height=7cm]{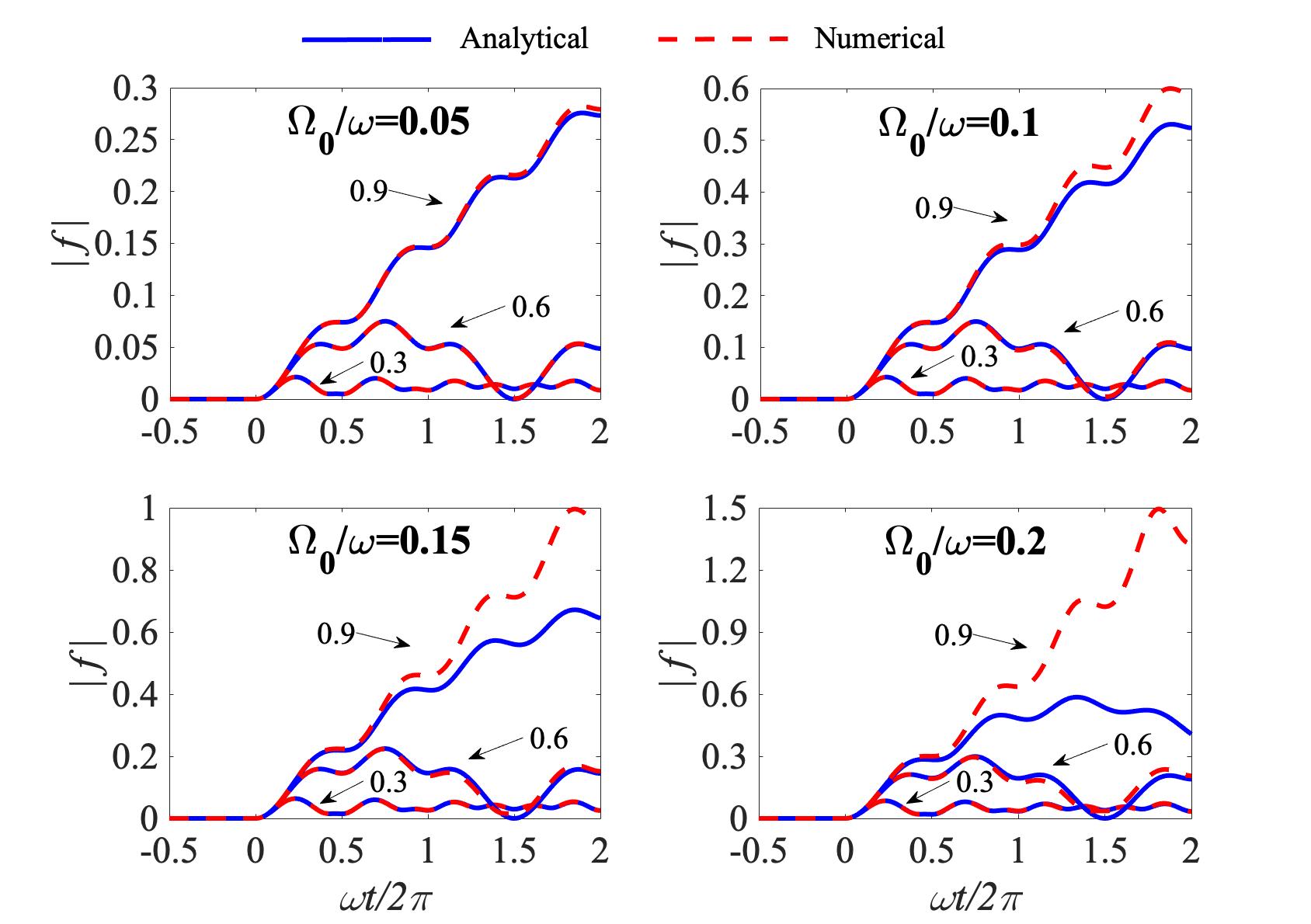} 
\captionsetup{justification=raggedright,singlelinecheck=false}
\caption{The analytical-numerical comparisons of $|f(t)|$ across the excitation pulse under four different peak fields, with $\Omega_0/\omega =$ 0.05, 0.1, 0.15 and 0.2. Three different detuning levels, with $\omega/\omega_c =$ 0.3, 0.6, and 0.9, are plotted in each case.}
\end{figure}

%%%%%%%%%%%%%%%%%%%%%%%Figure 3
\begin{figure}
\centering
\includegraphics[width=8.6cm, height=7cm]{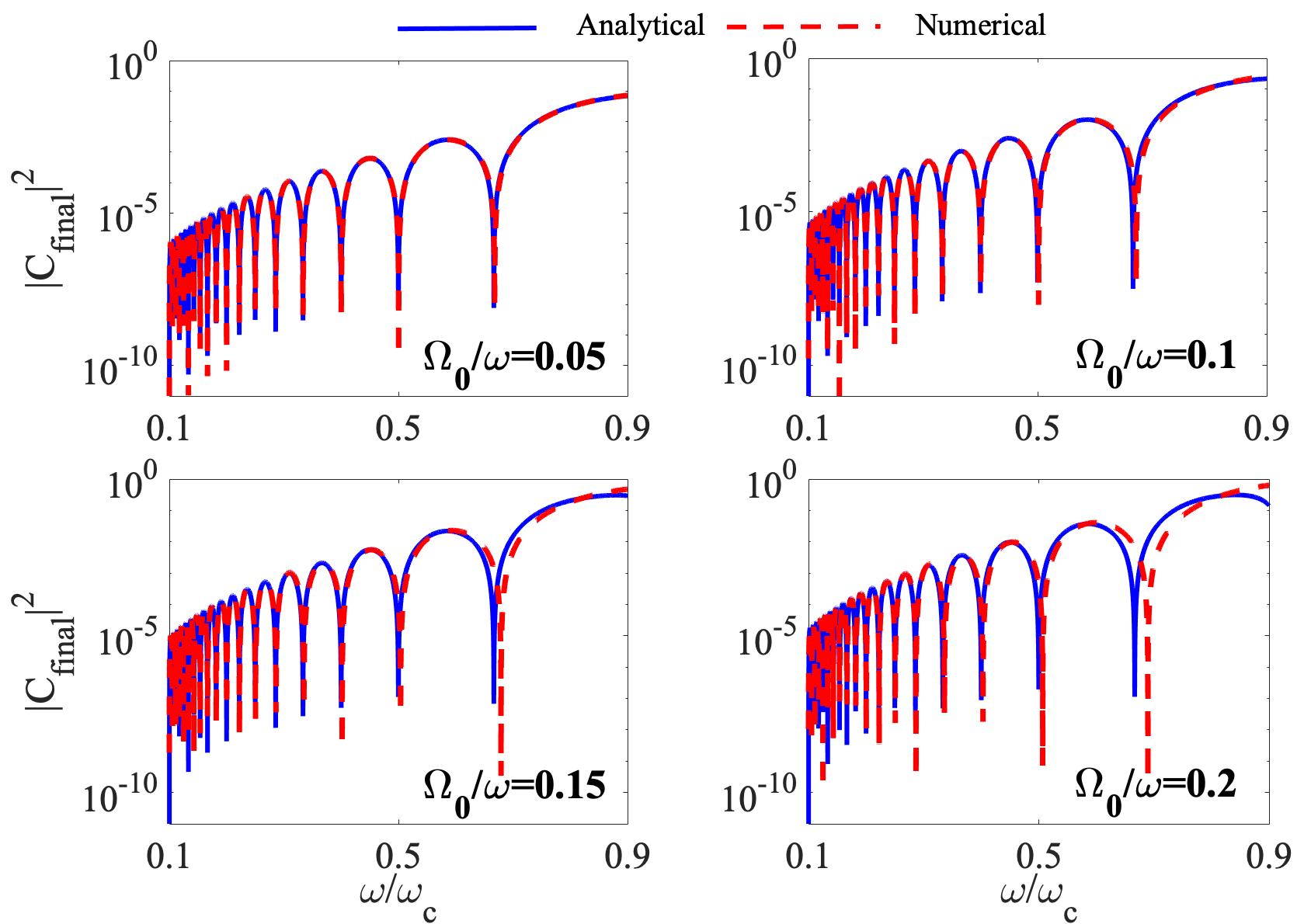} 
\captionsetup{justification=raggedright,singlelinecheck=false}
\caption{The final upper-level probability $|$C\textsubscript{final}$|^2$ versus the detuning factor $\omega/\omega_c$ under different excitation levels, with $\Omega_0/\omega =$ 0.05, 0.1, 0.15 and 0.2, respectively. Good analytical-numerical agreement is found in general except for the cases with large $\omega/\omega_c$ (small detuning).}
\end{figure}

Similar comparisons can also be made with respect to the final upper-level probability $|$C\textsubscript{final}$|^2$ at the end of the pulse, where $|$C$(t)|^2$ is related to $f(t)$ through the relation $|$C$(t)|^2 = |f(t)|^2/(1+|f(t)|^2)$. In Fig. 3, $|$C\textsubscript{final}$|^2$ is plotted against $\omega/\omega_c$ for the cases of $\Omega_0/\omega =$ 0.05, 0.1, 0.15 and 0.2. Once again, excellent analytical-numerical agreement is achieved with small $\Omega_0/\omega$ across a wide detuning range, whereas increasing disagreement is observed at small detunings as $\Omega_0$ approaches $\omega$, as evident from the $\Omega_0/\omega = 0.15$ and $\Omega_0/\omega = 0.2$ traces.

%%%%%%%%%%%%%%%%%%%%%%%Figure 4
\begin{figure}
\centering
\begin{subfigure}[b]{0.5\textwidth}
\includegraphics[width=7.4cm, height=4.5cm]{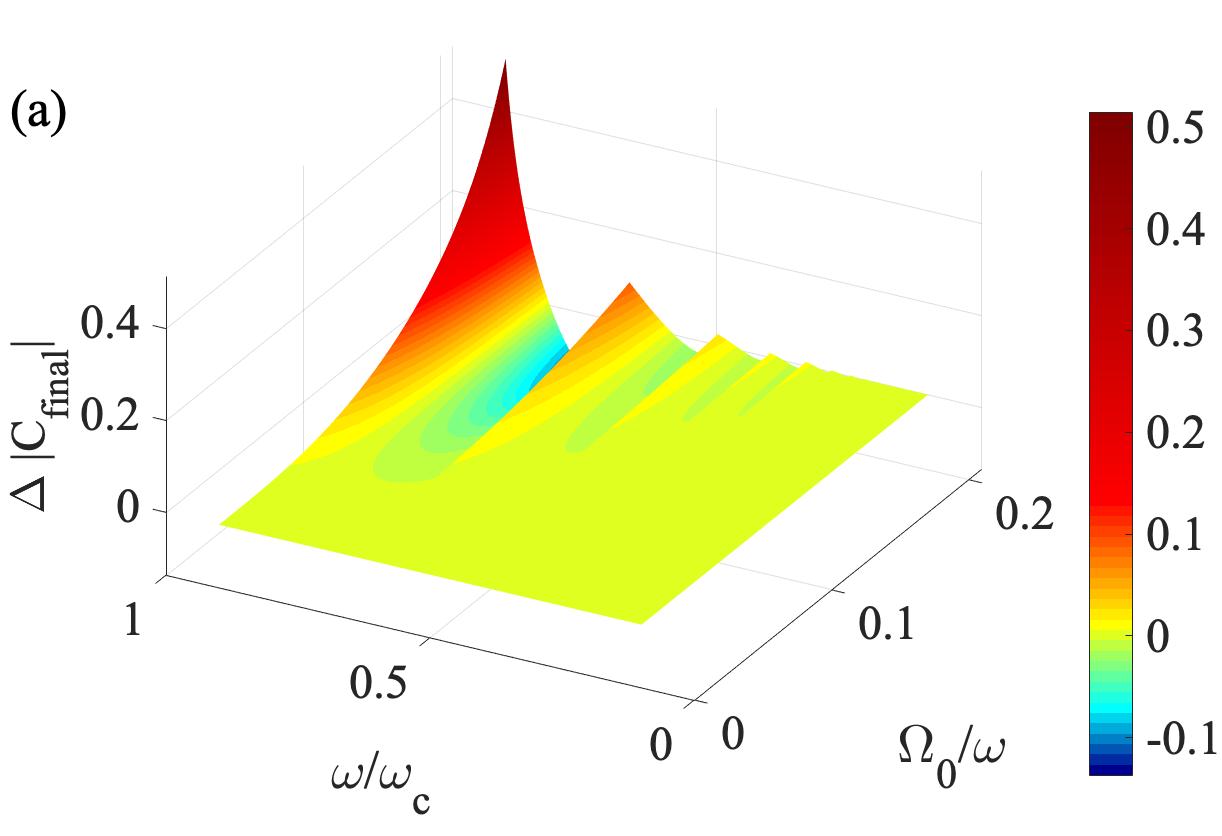} 
\end{subfigure}
\begin{subfigure}[b]{0.5\textwidth}
\includegraphics[width=7.4cm, height=4.5cm]{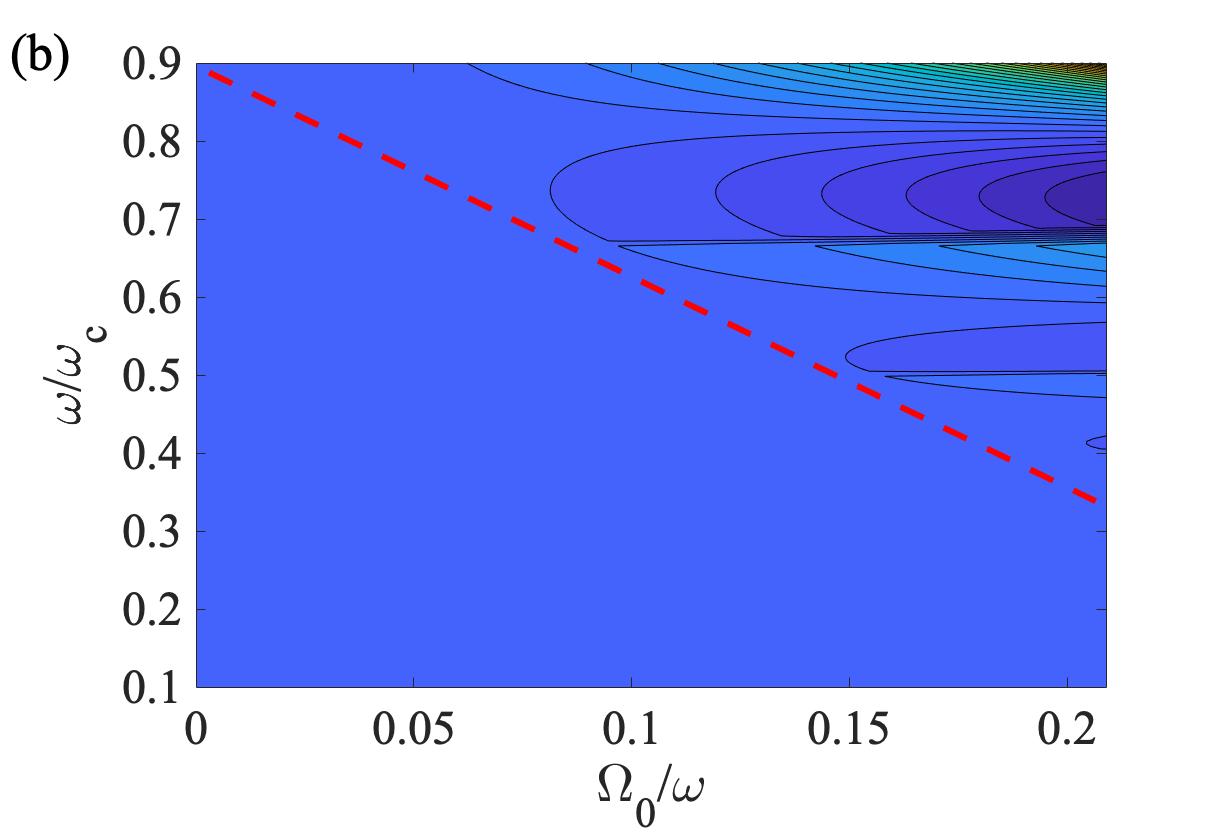} 
\end{subfigure}
\captionsetup{justification=raggedright,singlelinecheck=false}
\caption{The agreement between the analytical and the numerical solutions of $|$C\textsubscript{final}$|$ vs. $\omega/\omega_c$ and $\Omega_0/\omega$: (a) surface plot; (b) contour plot. The dashed guideline in (b) highlights the valid condition given by (16) for the analytical solution.}
\end{figure}

The above comparisons suggest that the analytical solution (16) is a valid solution for the Schrödinger equation (2) under the conditions of far off-resonance excitation and a \textit{small} peak incident field, where "small" is in reference to the threshold of carrier-wave Rabi flopping. To further verify this conclusion, the difference between the analytical and the numerical results, $\Delta|$C\textsubscript{final}$|$, is plotted on the two-dimensional space formed by $\Omega_0/\omega$ and $\omega/\omega_c$. The results are summarized by the surface plot shown in Fig. 4(a) and, equivalently, by the contour plot shown in Fig. 4(b). A diagonal guideline is added to Fig. 4(b) to highlight the boundary between the area where $\Delta|$C\textsubscript{final}$|=0$ and the area where nonzero $\Delta|$C\textsubscript{final}$|$ is generally observed. It is easy to see that the resulted valid condition qualitatively agrees with the theoretical prediction given by (17).

\section{CEP-Dependence of the Final Inversion}

The analytical solution given by (16) offers some interesting physical insights that numerical solutions would not be able to provide. For example, many prior studies have pointed out that the inversion of a TLS can be controlled by the CEP of a few-cycle pulse \cite{Jirauschek2005,Roudnev2007,Zeng2018}. This bears some practical implications given that the inversion is a direct measurable and hence can be used for experimental verification of quantum coherence. However, such a dependence has never been explicitly shown in closed form. The current theory has made it possible to derive such a relation.

We begin by recognizing that the inversion $w$ of a TLS can be expressed as $w = |C|^2-|D|^2 = (|f|^2-1)/(|f|^2+1)$, where $|f|^2$ can be derived from (16),
\begin{equation}
|f|^2 = \frac{1}{2} \left( 1 + \cos{\eta^2\omega_c t} \right) |\theta|^2.
\end{equation}

\noindent Using (12a) and (12b), it is straightforward to express $|\theta|^2$ as
\begin{equation}
\begin{split}
|\theta|^2 \equiv \theta \theta^* = {}& \eta^2 \{ \cosh{2\beta} \cdot (1 - \cos{\omega t} \cos{\omega_c t}) \\
& + \cos{(\omega t + 2\varphi)} (\cos{\omega t} - \cos{\omega_c t}) \\
& - \sinh{2\beta} \cdot \sin{\omega t} \cdot \sin{\omega_c t} \}.
\end{split}
\end{equation}

\noindent Typically, we are interested in the \textit{final} inversion of the TLS at the end of the driving pulse. For simplicity, we further assume that the square pulse contains an integer number of optical cycles. We then replace all the $t$ in (19) by $2N\pi/\omega$, which leads to
\begin{equation}
|\theta_f|^2 = 2 \eta^2 \cos^2{\varphi} \cdot \left( 1 - \cos{2 N \frac{\omega_c}{\omega} \pi} \right),
\end{equation}

\noindent where $N$ is a nonzero integer and the subscription $f$ indicates the final state upon the passage of the driving pulse. Note that the relation $\cosh{2\beta} = (\omega_c^2+\omega^2)/(\omega_c^2-\omega^2) \approx 1$ has been used in (20) under the far-detuning condition $\omega \ll \omega_c$. Substituting (20) into (18) and applying the definition of $w$ yield a concise expression of the final inversion
\begin{equation}
w_f = \frac{Q \eta^2 \cos^2{\varphi} - 1}{Q \eta^2 \cos^2{\varphi} + 1},
\end{equation}

\noindent where 
\begin{equation}
Q = \left( 1 + \cos{2 N \eta^2 \frac{\omega_c}{\omega} \pi} \right) \left( 1 - \cos{2 N \frac{\omega_c}{\omega} \pi} \right).
\end{equation}

%%%%%%%%%%%%%%%%%%%%%%%Figure 5
\begin{figure}
\centering
\begin{subfigure}[b]{0.5\textwidth}
\includegraphics[width=7.4cm, height=4.5cm]{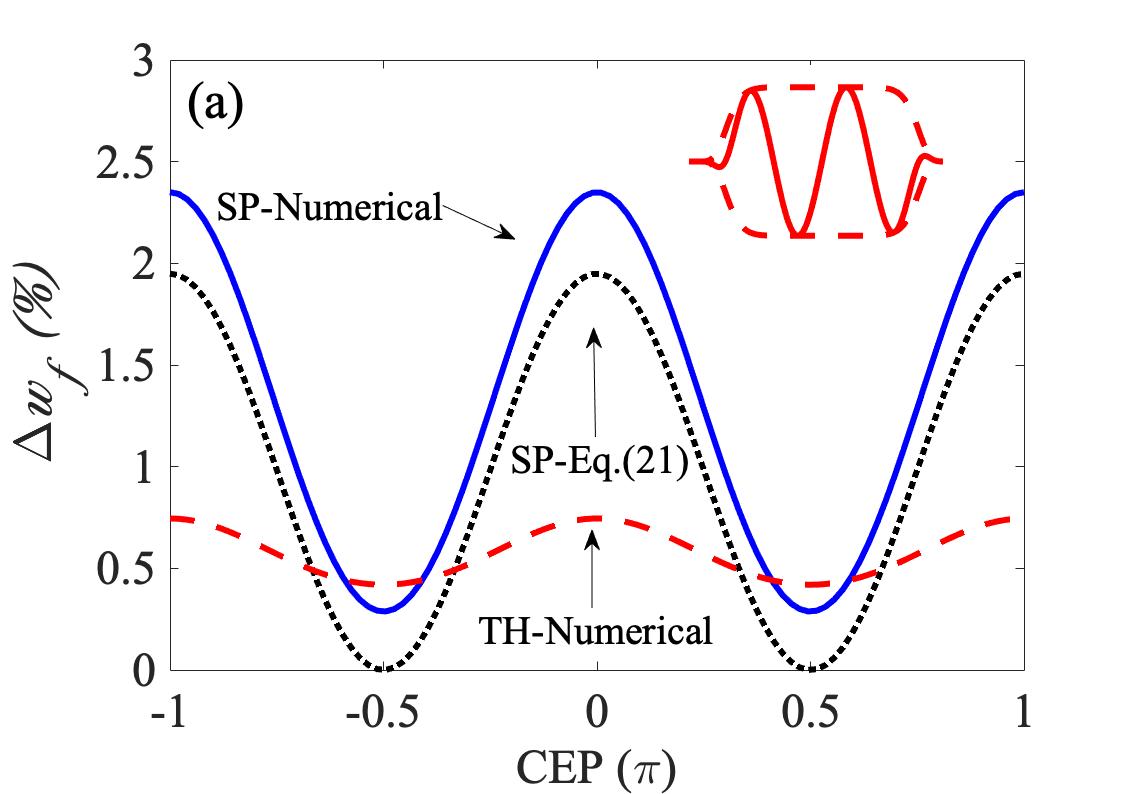} 
\end{subfigure}
\begin{subfigure}[b]{0.5\textwidth}
\includegraphics[width=7.4cm, height=4.8cm]{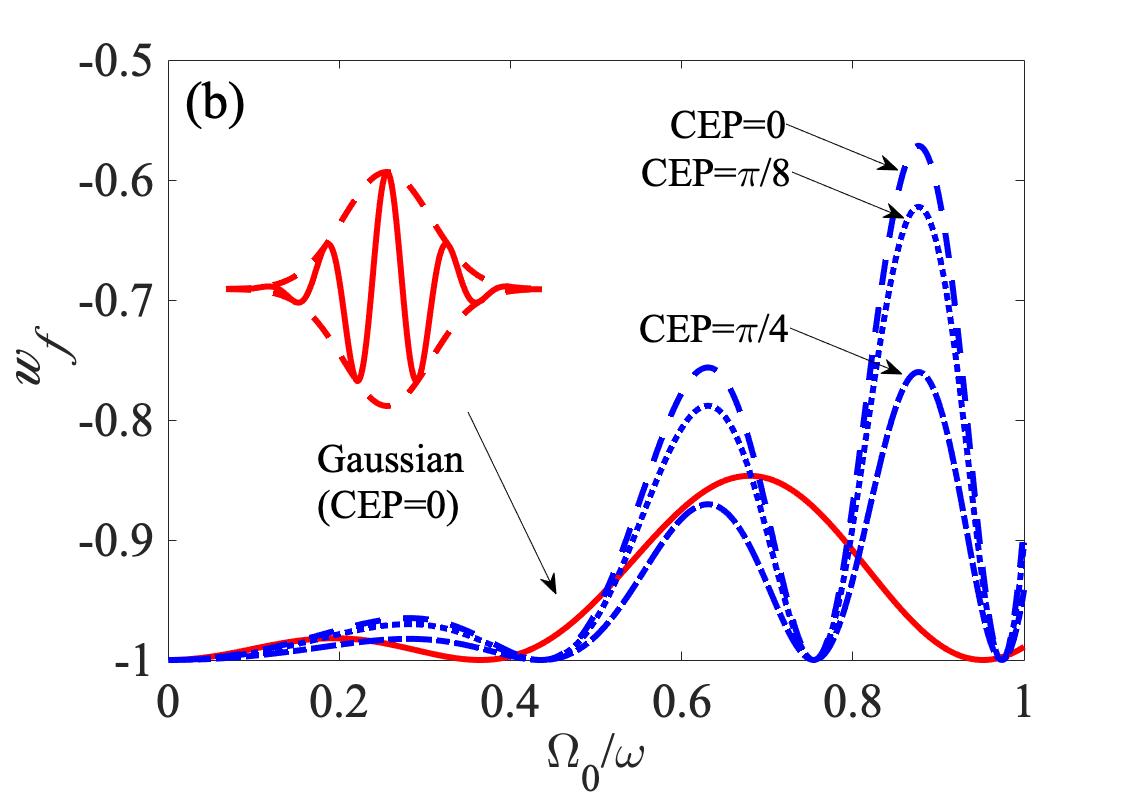} 
\end{subfigure}
\captionsetup{justification=raggedright,singlelinecheck=false}
\caption{(a) The CEP-dependence of the final inversion, where $\Delta w_f \equiv (w_f+1)/2$. The solid trace is obtained by numerically solving the Schrödinger equation for a square-pulse (SP), while the dotted trace is directly from the analytical solution (21). $\omega/\omega_c=0.366$ and $\Omega_0/\omega=0.181$ are used in both cases. The dashed trace is a numerical solution based on a two-cycle top-hat (TH) pulse (inset) with $\Omega_0/\omega \approx 0.28$. (b) The relation of $w_f$ vs. peak field $\Omega_0/\omega$, obtained from (23) for three different CEP values. The solid trace is a numerical solution based on a 1.5-cycle Gaussian pulse (inset) with a zero CEP.}
\end{figure}

Equation (21) highlights the direct impact of CEP on the inversion of a TLS. An immediate finding is that this CEP-dependence takes the form of $\cos^2{\varphi}$, which means $\varphi = 0$ and $\varphi = \pi$ result in the same inversion. In the current case, the minimum inversion occurs when $\varphi = \pm \pi/2$. These qualitative observations have been confirmed by comparing the numerical solution of the Schrödinger equation (2) with the results given by (21), as shown in Fig. 5(a) by the solid (numerical) and the dotted (analytical) traces. Aside from a small discrepancy in the absolute value, which results from the approximation of $\cosh{2\beta} \approx 1$, (21) offers a fairly accurate picture of the behaviors of the final inversion versus the CEP. Moreover, it is believed that these behaviors can be generalized to more realistic pulse shapes. As evidence, we show in Fig. 5(a) the numerical result based on a two-cycle top-hat pulse (depicted in the inset). The solution shares a similar CEP-dependence as the square pulse, albeit in a smaller variation scale.

Further insights can be obtained by using the fact that $Q \eta^2 \cos^2{\varphi}$ is typically much less than 1 and hence rewriting (21) as
\begin{equation}
w_f \approx -1 + 2Q \eta^2 \cos^2{\varphi}.
\end{equation}

\noindent Clearly, the effect of the CEP on $w_f$ is modulated by $\eta^2$ and, more importantly, by $Q$, which is given by (22). The value of $Q$ is bounded within $0 \leq Q \leq 4$, with $Q = 0$ occurring periodically when $\omega_c / \omega$ is an integer or $\eta^2 \omega_c / \omega$ is a half integer. This leads to an important realization that, when seeking an CEP-dependent inversion, it is critical to properly choose $\omega$, $\omega_c$ and $\Omega_0$ to maximize $Q$. On the other hand, the final inversion is locked to $-1$ when $Q = 0$, regardless of the CEP value. This is especially interesting from the view point of $\eta^2 \omega_c / \omega$ (the first term in (22)), as it suggests that certain peak-field values can override the effect of the CEP. Such a phenomena has been observed in numerical studies in the past, but has not been given a clear explanation before \cite{Roudnev2007,Zeng2020}. To further elucidate this point, Fig. 5(b) shows the dependence of the final inversion over the peak field of the square pulse for three different CEP values. Despite their different oscillation amplitudes, all three traces reach $w_f = -1$ at the same values of $\Omega_0/\omega$. Once again, these conclusions are not restricted to square-pulse excitation. For instance, we have numerically calculated the CEP-dependence of $w_f$ with various $\Omega_0$ for a 1.5-cycle Gaussian pulse and found similar characteristics. One of these traces with CEP $=0$ is included in Fig. 5(b). The trace shares the same features as the square-pulse traces with an oscillatory behavior and multiple minima at $w_f = -1$.

\section{Conclusion}

In conclusion, a closed-form analytical solution of the Schrödinger equation has been derived to describe the excitation of a TLS by a far-detuned, few-cycle square pulse without invoking the RWA or the SVEA. Despite its mathematical simplicity, the solution appears to capture some of the key aspects of the interaction, including its dependence on the CEP, and has shown very good agreements with numerical solutions. Some of the predictions of the theory prove to be robust against change of pulse shapes, demonstrating the potential of this analytical formalism as a generic description of light-matter interactions in the few-cycle regime. It is hoped that this work can shed some light on the path towards highly efficient control of quantum coherence along with other relevant topics.

\section{Funding}
This work has received funding from the National Science Foundation (NSF) (ECCS-1254902 and ECCS-1606836).

\nocite{*}

\bibliography{reference}% Produces the bibliography via BibTeX.

\end{document}